\documentclass[aps,pra,showpacs]{revtex4}
\usepackage{graphicx}

\begin{document}

\title{Generation of a highly phase sensitive polarization
squeezed $N$-photon state by collinear
parametric downconversion and coherent photon subtraction}

\author{Holger F. Hofmann}
\email{h.hofmann@osa.org}
\affiliation{
Graduate School of Advanced Sciences of Matter, Hiroshima University,
Kagamiyama 1-3-1, Higashi Hiroshima 739-8530, Japan}

\begin{abstract}
It is shown that a highly phase sensitive polarization
squeezed $(2n\!-\!1)$-photon state can be generated by
subtracting a diagonally polarized photon from the $2n$
photon component generated in collinear type II
downconversion. This polarization wedge state
has the interesting property that its photon number
distribution in the horizontal and vertical polarizations
remains sharply defined for phase shifts
of up to $1/n$ between the circularly polarized components.
Phase shifts at the Heisenberg limit are therefore observed
as nearly deterministic transfers of a single photon between
the horizontal and vertical polarization components.
\end{abstract}

%\keywords{}

\maketitle

\section{Introduction}

One of the most fundamental applications of non-classical
light field states is the improvement of measurement
precision beyond the standard quantum limits
for classical light sources. Of particular interest
is the possible enhancement of phase sensitivity in
interferometry \cite{Hol93,San95,Ou97,Ber00,Sod03,Wan05,Com05},
which could be useful in a wide range of fields, from
quantum lithography \cite{Jac95,Fon99,Bot00,Ang01}
to atomic clocks \cite{Chu00,Jos00}.
It is well known that the optimal phase resolution
$\Delta \Phi$ that can be achieved using a non-classical
$N$-photon state is given by the Heisenberg limit of
$\Delta \Phi \geq 1/N$. Recently, few photon interferometry
at this limit has been accomplished by new methods of generating
$N$-photon path entangled states using parametric
downconversion and post-selection \cite{Eda02,Fiu02,Hof04,Mit04,Eis05}.
Such path entangled states are an equal superposition
of the two $N$-photon states where all photons are located
in the same optical mode,
$(\mid N;0 \rangle + \mid 0,N \rangle)/\sqrt{2}$.
They are therefore ideally suited to obtain $N$-photon
interference fringes with a period of $2 \pi/N$ in
the optical phase shift between the two paths.
In principle, the generation of path entangled states can
be extended to higher photon numbers using the methods
proposed and realized in \cite{Fiu02,Hof04,Mit04,Eis05}.
In practice, however, the statistical bottlenecks
in the post-selection (or heralding) used to generate the
path entangled states rapidly reduce the
probabilities of generating an appropriate output as
photon number increases. It may therefore be useful to
consider alternative few photon states that can be generated
more efficiently from a given number of downconverted photon
pairs.

%%%
In this paper, it is shown that a highly phase
sensitive state can be generated by subtracting
a single diagonally polarized photon from the $(2n)$-photon
state generated in collinear type II downconversion.
Since single photon subtraction can be performed with
equal efficiency for any number of input photons, this
method could be very helpful in achieving phase resolutions
at the Heisenberg limit for higher photon numbers.
Moreover, the coherence induced between two adjacent
photon number states ensures that the narrowness of the
photon number distribution is maintained under phase
shifts of up to $1/n$. Phase shifts at the Heisenberg limit
can therefore be observed as nearly deterministic transfers
of a single photon between the output modes.

\section{Generation of the wedge state superposition}

\begin{figure}[t]
\begin{picture}(400,220)
%%\put(0,0){\framebox(400,220){}}
\thicklines
\put(20,122){\framebox(60,36){\Large PDC}}
%%--light beams
\put(80,137){\line(1,0){150}}
\put(80,143){\line(1,0){150}}
\put(100,115){\makebox(60,20){\large $\mid n;n \rangle_{HV}$}}
\put(178,137){\line(0,-1){62}}
\put(172,143){\line(0,-1){68}}
\put(175,72){\line(1,1){10}}
\put(175,72){\line(-1,1){10}}
\put(120,45){\makebox(80,20){\large $\frac{1}{\sqrt{2}}
\big(\mid\! n;n\!-\!1 \rangle_{HV}$}}
\put(150,25){\makebox(80,20){\large $ +\;
\mid\! n\!-\!1;n \rangle_{HV}\big)$}}
\put(235,137){\line(1,0){35}}
\put(235,143){\line(1,0){35}}
\put(300,137){\line(1,0){30}}
\put(300,143){\line(1,0){30}}
\put(333,140){\line(-1,1){10}}
\put(333,140){\line(-1,-1){10}}
\put(288,125){\line(0,-1){30}}
\put(282,125){\line(0,-1){30}}
\put(285,92){\line(1,1){10}}
\put(285,92){\line(-1,1){10}}
%%--Beam Splitter
\put(150,175){\makebox(50,15){\large Beam}}
\put(150,160){\makebox(50,15){\large splitter}}
\put(160,155){\line(1,-1){30}}
\put(190,110){\makebox(15,15){\large $R$}}
%%--halfwaveplate
\put(220,175){\makebox(25,15){\large $\lambda/2$-}}
\put(220,160){\makebox(25,15){\large plate}}
\put(230,125){\framebox(5,30){}}
\put(220,105){\makebox(25,15){$(22.5^\circ)$}}
%%--PBS
\put(270,160){\makebox(30,15){\large PBS}}
\put(270,155){\line(1,-1){30}}
\put(270,125){\framebox(30,30){}}
%%--Detectors
\put(290,100){\makebox(80,15){\large Photon}}
\put(290,85){\makebox(80,15){\large detectors}}
\put(340,130){\line(0,1){20}}
\bezier{100}(340,130)(350,130)(350,140)
\bezier{100}(340,150)(350,150)(350,140)
\bezier{100}(350,140)(360,140)(360,130)
\put(350,110){\makebox(20,20){\Large ``1"}}
\put(275,85){\line(1,0){20}}
\bezier{100}(275,85)(275,75)(285,75)
\bezier{100}(295,85)(295,75)(285,75)
\bezier{100}(285,75)(285,65)(295,65)
\put(298,55){\makebox(20,20){\Large ``0"}}
\end{picture}
\caption{\label{setup} Sketch of the experimental setup
generating the highly phase sensitive polarization wedge state.
The $2n\!-\!1$ photon state is generated by parametric downconversion
(PDC) of $n$ photon pairs, followed by a reflection of $2n\!-\!1$
photons at a beam splitter of reflectivity $R$ and detection
of the transmitted photon in a diagonally polarized state using
a $\lambda/2$-plate set at $22.5^\circ$ and a polarization
beam splitter (PBS).}
\end{figure}
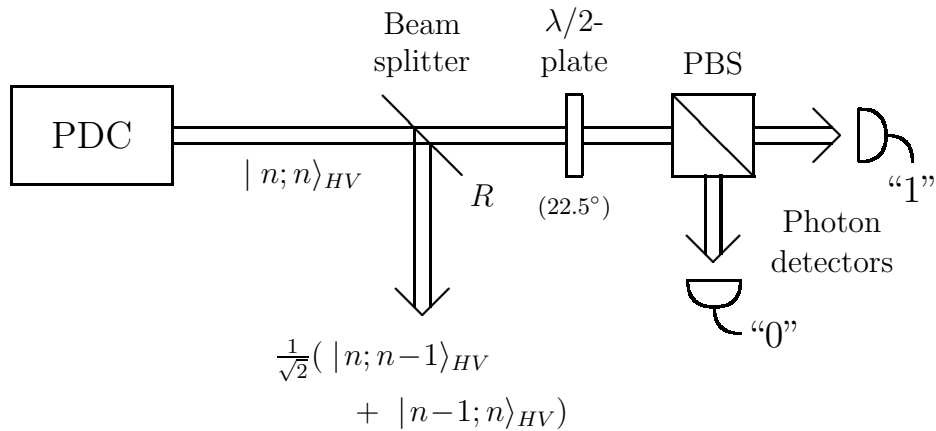

The proposed experimental setup is shown schematically
in fig. \ref{setup}. The initial state generated by collinear
type II parametric downconversion is a superposition of
photon number states with equal photon number in the horizontal
and vertical polarizations,
\begin{equation}
\label{eq:pdc}
\mid \mbox{PDC} \rangle =
\frac{1}{\cosh r} \sum_{n=0}^\infty (\tanh r)^n
\mid \! n; n \rangle_{HV}.
\end{equation}
If it can be assumed that all of the emitted photons will
eventually be detected, it is possible to isolate a single
$2n$-photon component by post-selecting only outputs where
a total of $2n$ photons are detected \cite{ps}.
Effectively, the input state is then given by
$\mid n; n \rangle_{HV}$. This $2n$-photon input
component is reflected at a beam splitter with a
reflectivity of $R$ close to one,
and one photon is detected in the transmitted light.
The components of the $2n$-photon states in the beam splitter
output with exactly one transmitted photon are given by
\begin{eqnarray}
\label{eq:bs}
\lefteqn{
\hat{U}_{R} \mid n; n \rangle_{HV} \otimes \mid 0; 0 \rangle_{HV}
\approx}
\nonumber \\[0.2cm] &&
\sqrt{n(1-R)R^{2n-1}}\left(
\mid \! n; n\!-\!1 \rangle_{HV} \otimes \mid 0; 1 \rangle_{HV}
+\mid \! n\!-\!1; n \rangle_{HV} \otimes \mid 1; 0 \rangle_{HV}
\right) + \ldots,
\end{eqnarray}
The beam splitter thus entangles the polarization of the
transmitted one photon component
and the polarization of the reflected $(2n-1)$-photon component.
It is now possible to measure the diagonal polarization of the
transmitted photon using a $\lambda/2$-plate set at $22.5^\circ$
and a polarization beam splitter. This measurement projects the
state of the transmitted photon onto an equal superposition
of horizontal and vertical polarization, resulting in a conditional
output state of
\begin{equation}
\label{eq:wedge}
\mid \mbox{Wedge} \rangle = \frac{1}{\sqrt{2}}
\left(\mid n; n\!-\!1 \rangle_{HV} + \mid n\!-\!1 ; n \rangle_{HV} \right)
\end{equation}
in the reflected light.

Equation (\ref{eq:bs}) shows that the probability of successfully
subtracting exactly one photon from the $(2n)$-photon input is
given by $n(1-R)R^{2n-1}$. This value can be optimized independently
for any desired photon number by varying the reflectivity $R$.
The maximal efficiency of photon subtraction is obtained at
$R=1-1/(2n)$. The probability of successful photon subtraction is
then equal to $(1-1/(2n))^{2n-1}/2$.
Interestingly, this maximal probability decreases
only slightly with photon number, from an initial value of 25\%
at $n=1$ towards a value of $1/(2e)\approx 18.4 \%$ for
extremely high photon numbers.
By selecting an optimized reflectivity of $R=1-1/(2n)$, it is
thus possible to achieve post-selection probabilities greater
than 18\% for any number of input photons.
The efficiency of photon subtraction is therefore almost
independent of photon number.

It should be noted that this is quite different from the photon
bottleneck used to generate the path entangled state
$(\mid N;0 \rangle + \mid 0,N \rangle)/\sqrt{2}$, where the
corresponding post-selection probability drops rapidly with
increasing photon number as more and more beam splitters become
necessary to "bunch up" the photons in the single mode bottleneck.
In the basic scheme introduced in \cite{Hof04}, the bottleneck
efficiency is $2 N!/(2 N)^N$ for an $N$ photon state. At 5 photons,
this is an efficiency of only 0.24 \%, almost a hundred times less
than the optimal efficiencies of photon subtraction.
It should therefore be much easier to increase the output photon
number of wedge states than to achieve the same photon number
for path entangled states.

\section{Stokes parameter statistics of ($2n\!-\!1$)-photon
wedge states}
\label{sec:stokes}

The complete polarization statistics of $N$-photon quantum
states can be expressed by the three Stokes parameters
describing the photon number differences between horizontal (H)
and vertical (V), plus (P) and minus (M) diagonal, and
right (R) and left (L) circular polarization,
\begin{equation}
\begin{array}{rcccc}
\hat{S}_1 &=& \hat{n}_H -\hat{n}_V &=&
\hat{a}_H^\dagger \hat{a}_H - \hat{a}_V^\dagger \hat{a}_V
\\
\hat{S}_2 &=& \hat{n}_P -\hat{n}_M &=&
\hat{a}_H^\dagger \hat{a}_V + \hat{a}_V^\dagger \hat{a}_H
\\
\hat{S}_3 &=& \hat{n}_R -\hat{n}_L &=&
-i (\hat{a}_H^\dagger \hat{a}_V - \hat{a}_V^\dagger \hat{a}_H).
\end{array}
\end{equation}
As can be seen from eq. (\ref{eq:wedge}), the Stokes parameter
$\hat{S}_1$ describing the $HV$-polarization takes on values
of $+1$ or $-1$, with a 50\% probability each. The average
of $\hat{S}_1$ is therefore zero, and its uncertainty is
$\delta S_1^2=1$.

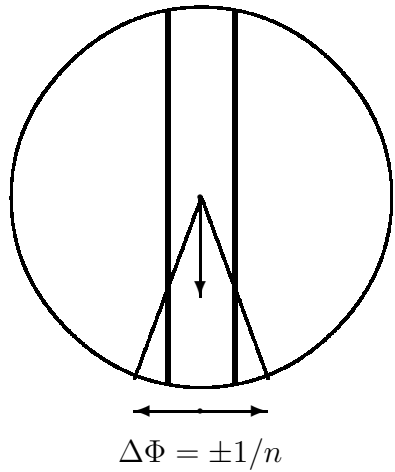
\begin{figure}[t]
\setlength{\unitlength}{1.8pt}
\begin{picture}(120,120)
%%\put(0,0){\framebox(240,200){}}
\thicklines
%%--distribution

%%--Poincare sphere
\bezier{200}(20,60)(20,76)(32,88)
\bezier{200}(32,88)(44,100)(60,100)
\bezier{200}(60,100)(76,100)(88,88)
\bezier{200}(88,88)(100,76)(100,60)
\bezier{200}(100,60)(100,44)(88,32)
\bezier{200}(88,32)(76,20)(60,20)
\bezier{200}(60,20)(44,20)(32,32)
\bezier{200}(32,32)(20,44)(20,60)

%\thinlines
%\put(25,101){\line(0,1){38}}
%\put(39,86){\line(0,1){68}}
%\put(53,81){\line(0,1){78}}
%\put(67,81){\line(0,1){78}}
%\put(81,86){\line(0,1){68}}
%\put(95,101){\line(0,1){38}}

%\thicklines
\put(53,21){\line(0,1){78}}
\put(53.5,21){\line(0,1){78}}
\put(67,21){\line(0,1){78}}
\put(67.5,21){\line(0,1){78}}
\put(60,60){\circle*{1}}
\put(60,60){\vector(0,-1){21}}

\bezier{200}(60,60)(53,41)(46,22)
\bezier{200}(60,60)(67,41)(74,22)

\put(60,15){\vector(-1,0){14}}
\put(60,15){\vector(1,0){14}}
\put(60,15){\circle*{1}}

\put(30,0){\makebox(60,12){\large $\Delta \Phi = \pm 1/n$}}

\end{picture}
\setlength{\unitlength}{1pt}
\caption{\label{wedge}
Schematic illustration of the wedge state statistics in the $S_1$-$S_2$
plane of the Poincare sphere. The scale chosen corresponds to five
photons ($n=3$). The arrow indicates the average Stokes vector,
the thick vertical lines indicate the quantized eigenstates of
$\hat{S}_1$. The thin lines at angles of $\pm \Delta \Phi$ illustrate
the phase uncertainty of the wedge state. As photon number increases,
the eigenstates with $S_1=\pm 1$ move closer together and the
phase distribution of the wedge state becomes narrower.
}
\end{figure}

The low uncertainty in $\hat{S}_1$ is a direct consequence
of the quantum correlations in parametric downconversion.
In fact, the original $\mid n; n\rangle_{HV}$-state is an
$\hat{S}_1$ eigenstate with an uncertainty of zero, which
already provides phase sensitivities at the Heisenberg limit
in the photon statistics \cite{Hol93}. Photon subtraction
actually increases the $\hat{S}_1$-uncertainty by one,
thus increasing the observed photon number noise in the $HV$-basis.
However, the essential effect of the photon subtraction on
the polarization statistics of the output state is the
generation of coherence between the horizontal and
vertical polarization components. This effect can be observed
in the statistics of the Stokes parameter $\hat{S}_2$, describing
the photon number difference between the diagonal polarizations
$P$ and $M$. Due to the coherence between $\mid n; n-1\rangle_{HV}$
and $\mid n-1; n\rangle_{HV}$, the expectation value of this
Stokes parameter is
\begin{equation}
\langle \hat{S}_2 \rangle = \langle \hat{n}_P - \hat{n}_M \rangle = n.
\end{equation}
Since the
total photon number is $N=2n\!-\!1$, this means that on average,
more than 3/4 of all output photons are polarized along the
same diagonal as the transmitted photon \cite{note}.
Since the wedge state polarization of $\langle \hat{S}_2 \rangle = n$
originates from complete coherence between two adjacent eigenstates
of $\hat{S}_1$, it is the maximal diagonal polarization possible
at an uncertainty of only $\delta S_1^2=1$ in the difference between horizontally and vertically polarized photons. As shown in fig.
\ref{wedge}, the polarization distribution of the output state thus resembles a quantum mechanically narrow "wedge" inserted between the $\hat{S}_1$ eigenstates from the positive side of the diagonal
polarization $\hat{S}_2$ - somewhat like the slice of an orange,
with a quantum limited thickness of $2 \delta S_1 =2$.

Due to the narrowness of its $\hat{S}_1$ distribution and due
to its comparatively high expectation value
$\langle \hat{S}_2 \rangle$, the wedge state is very suitable for
measurements of small phase shifts between the right and left
circular polarizations, which result in a rotation of the Stokes
parameters around the $S_3$ axis. As indicated in fig. \ref{wedge}, the
phase uncertainty of the polarization wedge can then be estimated by
\begin{equation}
\label{eq:delta}
\delta \Phi^2 = \frac{\delta S_1^2}{\langle S_2 \rangle^2} =
\frac{1}{n^2}.
\end{equation}
This phase uncertainty is close to the Heisenberg limit and
corresponds to the phase uncertainties of the various phase
squeezed states proposed for optimized phase estimation in
quantum interferometry \cite{Com05}.
The wedge state is therefore an almost ideal phase squeezed
$N$-photon state.

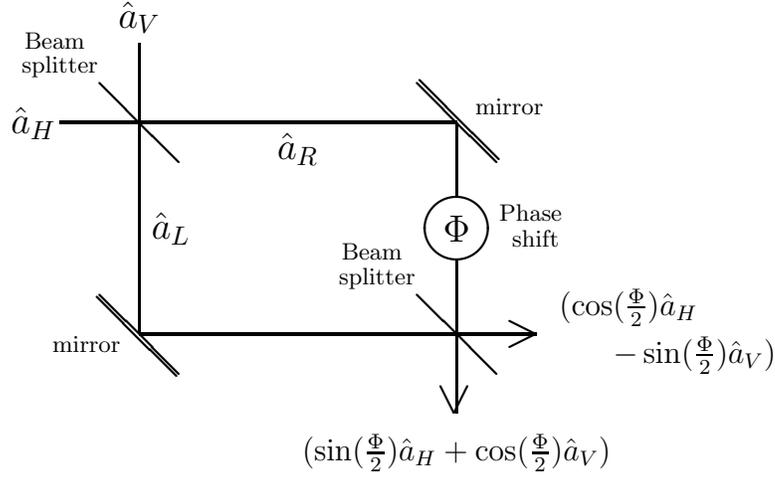
\begin{figure}
\begin{picture}(380,200)
%%\put(0,0){\framebox(380,200){}}
\thicklines
\put(70,130){\makebox(20,20){\Large $\hat{a}_H$}}
\put(90,140){\line(1,0){150}}
\put(110,170){\makebox(20,20){\Large $\hat{a}_V$}}
\put(120,170){\line(0,-1){110}}
\put(170,120){\makebox(20,20){\Large $\hat{a}_R$}}
\put(122,90){\makebox(20,20){\Large $\hat{a}_L$}}
\put(240,140){\line(0,-1){28}}
\put(240,100){\circle{24}}
\put(255,100){\makebox(26,12){Phase}}
\put(255,90){\makebox(30,12){shift}}
\put(225,85){\makebox(30,30){\Large $\Phi$}}
\put(240,88){\line(0,-1){58}}
\put(239,30){\line(-1,2){5}}
\put(239,30){\line(1,2){5}}
\put(120,60){\line(1,0){150}}
\put(270,60){\line(-2,1){10}}
\put(270,60){\line(-2,-1){10}}

\put(270,60){\makebox(70,20){\large $(\cos(\frac{\Phi}{2}) \hat{a}_H$}}
\put(295,42){\makebox(70,20){\large $- \sin(\frac{\Phi}{2}) \hat{a}_V)$}}

\put(180,5){\makebox(120,20){\large $(\sin(\frac{\Phi}{2}) \hat{a}_H
+ \cos(\frac{\Phi}{2}) \hat{a}_V)$}}

\put(70,165){\makebox(36,12){Beam}}
\put(70,155){\makebox(40,12){splitter}}
\put(105,155){\line(1,-1){30}}

\put(80,50){\makebox(40,12){mirror}}
\put(105,75){\line(1,-1){30}}
\put(104,74){\line(1,-1){30}}

\put(240,140){\makebox(40,12){mirror}}
\put(225,155){\line(1,-1){30}}
\put(226,156){\line(1,-1){30}}

\put(190,85){\makebox(36,12){Beam}}
\put(190,75){\makebox(40,12){splitter}}
\put(225,75){\line(1,-1){30}}

\end{picture}

\caption{\label{MZ}
Illustration of interferometry using the wedge state coherence.
By modifying the phases of the $RL$ modes, the original $HV$
input is rotated towards the $PM$ basis.
In polarization experiments, the same effect can be achieved
by a single half wave plate set at $\theta=\phi/4$ followed by
a polarization beam splitter.}
\end{figure}

To clarify the application of this phase sensitivity in
interferometry, it may be useful to consider the possibility
of converting the polarization modes $\hat{a}_H$ and $\hat{a}_V$
to spatial input modes with equal polarization.
As shown in fig. \ref{MZ}, the two paths inside
the interferometer then correspond to the modes
$\hat{a}_{R/L}=(\hat{a}_H\pm i\hat{a}_V)/\sqrt{2}$, and the
effect of a phase shift $\Phi$ between the two paths is to
rotate the original $HV$ basis towards the $PM$ basis.
In terms of the photon number difference $\hat{S}_1(\mbox{out})$
observed in the two output ports, this effect can then be expressed
as the rotation of the Stokes vector around the $S_3$ axis mentioned
above, with
\begin{equation}
\hat{S}_1 (\mbox{out})
= \cos[\Phi]\; \hat{S}_1 (\mbox{in})
+ \sin[\Phi]\; \hat{S}_2 (\mbox{in}).
\end{equation}
The phase shift $\Phi$ between the arms of the interferometer
thus corresponds directly to the rotation of the linear polarization
components obtained e.g. by a half wave plate set at $\theta=\phi/4$.
The phase sensitivity of the $(2n-1)$-photon wedge state
can then be described by the average
$\langle \hat{S}_1 (\mbox{out})\rangle$
and the variance $\delta S_1^2(\mbox{out})$ of the
measurement result $S_1(\mbox{out})=n_H-n_V$,
\begin{eqnarray}
\label{eq:stats}
\langle \hat{S}_1(\mbox{out})\rangle &=&
n \sin[\Phi]
\nonumber \\[0.2cm]
\delta S_1^2(\mbox{out}) &=& 1+(n^2-2) \sin^2[\Phi].
\end{eqnarray}
For phase shifts $\Phi$ smaller than $1/n$,
$\langle \hat{S}_1 \rangle \approx n \Phi$ and
$\delta S_1^2 \approx 1$ corresponds to a phase resolution
of $\Delta \Phi = 1/n$, as given by eq.(\ref{eq:delta})
above. ($2n-1$)-photon wedge states can thus
achieve phase resolutions close to the Heisenberg limit
for arbitrarily high photon numbers.
Moreover, the variance of the output photon number distribution at
phase shifts $\Phi$ with $n \sin[\Phi]=\pm 1$ is still smaller
than two. Even at phase shifts that change the average output
photon number difference by one,
the photon number distribution is therefore sharper than
the difference of two between two adjacent measurement
outcomes of $\hat{S}_1$. This result indicates that phase shifts
at the Heisenberg limit are observed as nearly deterministic
transfers of a single photon between the horizontal and
vertical polarizations, with measurement probabilities greater
than 50\% of finding a measurement outcome equal to the
expectation value of $\langle \hat{S}_1 \rangle = \pm 1$ at
$n \sin[\Phi]=\pm 1$.

It is interesting to note that the above argument only relies on
averages and variances of the photon number differences $\hat{S}_i$.
It is therefore straightforward to estimate the effect of
basic photon counting errors. In particular, a photon loss error
may occur when the downconversion actually generates $n+1$ pairs,
and two photons are subsequently lost due to limited
detector efficiencies. At low detector efficiencies, the
probability of such a photon loss error
is of the order of $(\tanh r)^2$, corresponding to the ratio of the
$n+1$ pair probability to the $n$ pair probability in eq.(\ref{eq:pdc}).
Since random (=unpolarized) photon losses do not change the
average polarization of the photons, the strong coherence between
the $H$ and $V$ polarization given by $\langle \hat{S}_2 \rangle$
is unchanged by this error. However, the polarization
fluctuations increase due
to the possibility that both of the photons lost had the same
polarization. Specifically, the increase in $\delta S_1^2$ is equal
to two times the probability of the photon loss error.
For reasonably low error probabilities
(e.g. for $(\tanh r)^2 < 0.1$),
this additional uncertainty is much smaller than the pure state uncertainty of $\delta S_1^2=1$, and the effect on the phase
resolution will be negligible. These considerations indicate
that the high phase resolution of wedge states is rather robust
against photon loss errors. Again, this is an important difference to
path entangled states, where the loss of a single photon completely
destroys the quantum coherence responsible for the high phase resolution.
Due to this robustness against photon loss errors, it may be
interesting to investigate wedge state generation at high pump
powers even if high detector efficiencies cannot
be achieved \cite{Eis04}.

\begin{figure}[t]
\begin{picture}(360,220)
%%\put(0,0){\framebox(360,220){}}
\put(80,195){\makebox(200,20){\large
Probability
$|\langle 2;3 \mid \hat{U}_\phi \mid \mbox{Wedge}\rangle|^2$
}}
\put(0,0){\makebox(320,220){
\scalebox{1.2}[1.2]{
\includegraphics{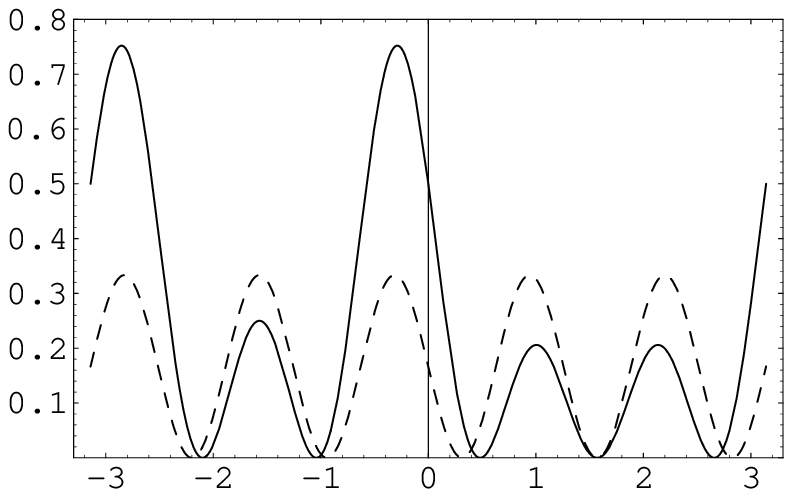}}
}}
\put(130,5){\makebox(100,20){\large
Phase shift $\Phi$
}}
\end{picture}
\caption{\label{fringes}
Probability of finding two photons in the horizontally polarized
mode and three photons in the vertically polarized mode as a
function of phase shift $\Phi$ between the circular polarizations
for the five photon wedge state.
The dotted line shows the same probability for a corresponding
five photon path-entangled state.
}
\end{figure}

\section{Photon statistics of the five photon wedge state}

To illustrate the full implications of the phase sensitivity of
wedge states at the level of precise photon counting,
it may be useful to take a closer look at a specific
example. Here, a compromise is necessary between the
experimental difficulties and the increasing phase resolution
permitted by higher photon numbers.
A good choice may be the five photon wedge state,
generated by subtracting one photon from $n=3$ pairs of
downconverted photons, since it should be just within reach
of present technological possibilities.
A phase shift of $\Phi$ transforms this state according to
\begin{eqnarray}
\langle 5;0 \mid \hat{U}_\phi \mid \mbox{Wedge} \rangle
&=& -\frac{\sqrt{10}}{16} \left(
    \cos[\frac{5}{2}\phi -\frac{\pi}{4}]
+   \cos[\frac{3}{2}\phi +\frac{\pi}{4}]
- 2 \cos[\frac{1}{2}\phi -\frac{\pi}{4}]\right)
\nonumber \\
\langle 4;1 \mid \hat{U}_\phi \mid \mbox{Wedge} \rangle
&=& -\frac{\sqrt{2}}{16} \left(
  5 \cos[\frac{5}{2}\phi +\frac{\pi}{4}]
- 3 \cos[\frac{3}{2}\phi -\frac{\pi}{4}]
- 2 \cos[\frac{1}{2}\phi +\frac{\pi}{4}]\right)
\nonumber \\
\langle 3;2 \mid \hat{U}_\phi \mid \mbox{Wedge} \rangle
&=& \frac{1}{8} \left(
  5 \cos[\frac{5}{2}\phi -\frac{\pi}{4}]
+   \cos[\frac{3}{2}\phi +\frac{\pi}{4}]
+ 2 \cos[\frac{1}{2}\phi -\frac{\pi}{4}]\right)
\nonumber \\
\langle 2;3 \mid \hat{U}_\phi \mid \mbox{Wedge}\rangle
&=& \frac{1}{8} \left(
  5 \cos[\frac{5}{2}\phi +\frac{\pi}{4}]
+   \cos[\frac{3}{2}\phi -\frac{\pi}{4}]
+ 2 \cos[\frac{1}{2}\phi +\frac{\pi}{4}]\right)
\nonumber \\
\langle 1;4 \mid \hat{U}_\phi \mid \mbox{Wedge} \rangle
&=& -\frac{\sqrt{2}}{16} \left(
  5 \cos[\frac{5}{2}\phi -\frac{\pi}{4}]
- 3 \cos[\frac{3}{2}\phi +\frac{\pi}{4}]
- 2 \cos[\frac{1}{2}\phi -\frac{\pi}{4}]\right)
\nonumber \\
\langle 5;0 \mid \hat{U}_\phi \mid \mbox{Wedge} \rangle
&=& -\frac{\sqrt{10}}{16} \left(
    \cos[\frac{5}{2}\phi +\frac{\pi}{4}]
+   \cos[\frac{3}{2}\phi -\frac{\pi}{4}]
- 2 \cos[\frac{1}{2}\phi +\frac{\pi}{4}]\right).
\end{eqnarray}
Each of these six amplitudes includes a five photon interference
component oscillating at a rate of $5\phi/2$. This five photon
interference effect is particularly strong in the $\mid 3;2 \rangle$
and the $\mid 2;3 \rangle$ components. Fig. \ref{fringes} shows
the interference fringes observed in the $\mid 2;3 \rangle$
component. For comparison, the dashed line shows the corresponding fringes of a five photon path-entangled state. The main difference
between the two fringes are the different peak heights of the
wedge state indicating the average diagonal polarization of the
five photon state. The wedge state thus combines features of the
maximal five photon interference of path-entangled states with
the well defined polarization direction of a phase squeezed state.

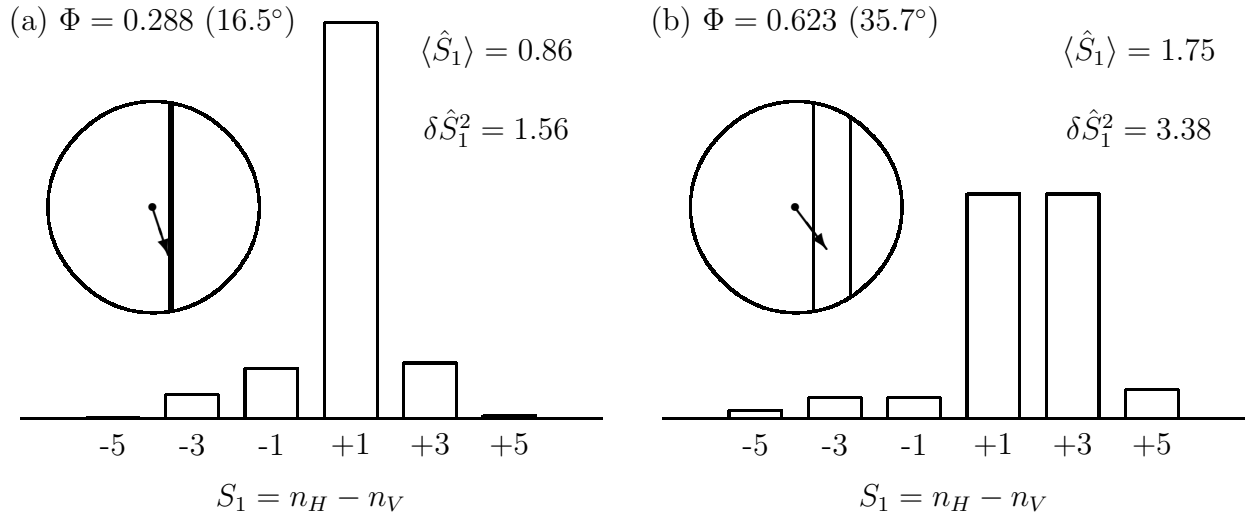
\begin{figure}[t]
\begin{picture}(240,220)
%%\put(0,0){\framebox(240,200){}}
\thicklines
\put(0,180){\makebox(120,20){\large (a) $\Phi=0.288$ ($16.5^\circ$)}}
%%--distribution
\put(10,40){\line(1,0){220}}
\put(35,40.5){\line(1,0){20}}
\put(35,20){\makebox(20,20){\large -5}}
\put(65,40){\line(0,1){9}}
\put(85,40){\line(0,1){9}}
\put(65,49){\line(1,0){20}}
\put(65,20){\makebox(20,20){\large -3}}
\put(95,40){\line(0,1){19}}
\put(115,40){\line(0,1){19}}
\put(95,59){\line(1,0){20}}
\put(95,20){\makebox(20,20){\large -1}}
\put(125,40){\line(0,1){150}}
\put(145,40){\line(0,1){150}}
\put(125,190){\line(1,0){20}}
\put(125,20){\makebox(20,20){\large +1}}
\put(155,40){\line(0,1){21}}
\put(175,40){\line(0,1){21}}
\put(155,61){\line(1,0){20}}
\put(155,20){\makebox(20,20){\large +3}}
\put(185,40){\line(0,1){1}}
\put(205,40){\line(0,1){1}}
\put(185,41){\line(1,0){20}}
\put(185,20){\makebox(20,20){\large +5}}
\put(80,0){\makebox(80,20){\large $S_1=n_H-n_V$}}

%%--Poincare sphere
\bezier{200}(20,120)(20,136)(32,148)
\bezier{200}(32,148)(44,160)(60,160)
\bezier{200}(60,160)(76,160)(88,148)
\bezier{200}(88,148)(100,136)(100,120)
\bezier{200}(100,120)(100,104)(88,92)
\bezier{200}(88,92)(76,80)(60,80)
\bezier{200}(60,80)(44,80)(32,92)
\bezier{200}(32,92)(20,104)(20,120)

%\thinlines
%\put(25,101){\line(0,1){38}}
%\put(39,86){\line(0,1){68}}
%\put(53,81){\line(0,1){78}}
%\put(67,81){\line(0,1){78}}
%\put(81,86){\line(0,1){68}}
%\put(95,101){\line(0,1){38}}

%\thicklines
\put(66.5,81){\line(0,1){78}}
\put(67.5,81){\line(0,1){78}}

\put(60,120){\circle*{3}}
\put(60,120){\vector(1,-3){6}}

\put(160,170){\makebox(60,20){\large $\langle \hat{S}_1\rangle = 0.86 $}}
\put(160,140){\makebox(60,20){\large $\delta \hat{S}_1^2 = 1.56 $}}

\end{picture}
\begin{picture}(240,220)
%%\put(0,0){\framebox(240,200){}}
\thicklines
\put(0,180){\makebox(120,20){\large (b) $\Phi=0.623$ ($35.7^\circ$)}}
%%--distribution
\put(10,40){\line(1,0){220}}
\put(35,40){\line(0,1){3}}
\put(55,40){\line(0,1){3}}
\put(35,43){\line(1,0){20}}
\put(35,20){\makebox(20,20){\large -5}}
\put(65,40){\line(0,1){8}}
\put(85,40){\line(0,1){8}}
\put(65,48){\line(1,0){20}}
\put(65,20){\makebox(20,20){\large -3}}
\put(95,40){\line(0,1){8}}
\put(115,40){\line(0,1){8}}
\put(95,48){\line(1,0){20}}
\put(95,20){\makebox(20,20){\large -1}}
\put(125,40){\line(0,1){85}}
\put(145,40){\line(0,1){85}}
\put(125,125){\line(1,0){20}}
\put(125,20){\makebox(20,20){\large +1}}
\put(155,40){\line(0,1){85}}
\put(175,40){\line(0,1){85}}
\put(155,125){\line(1,0){20}}
\put(155,20){\makebox(20,20){\large +3}}
\put(185,40){\line(0,1){11}}
\put(205,40){\line(0,1){11}}
\put(185,51){\line(1,0){20}}
\put(185,20){\makebox(20,20){\large +5}}
\put(80,0){\makebox(80,20){\large $S_1=n_H-n_V$}}

%%--Poincare sphere
\bezier{200}(20,120)(20,136)(32,148)
\bezier{200}(32,148)(44,160)(60,160)
\bezier{200}(60,160)(76,160)(88,148)
\bezier{200}(88,148)(100,136)(100,120)
\bezier{200}(100,120)(100,104)(88,92)
\bezier{200}(88,92)(76,80)(60,80)
\bezier{200}(60,80)(44,80)(32,92)
\bezier{200}(32,92)(20,104)(20,120)

%\thinlines
%\put(25,101){\line(0,1){38}}
%\put(39,86){\line(0,1){68}}
%\put(53,81){\line(0,1){78}}
%\put(67,81){\line(0,1){78}}
%\put(81,86){\line(0,1){68}}
%\put(95,101){\line(0,1){38}}

%\thicklines
\put(67,81){\line(0,1){78}}
\put(81,86){\line(0,1){68}}

\put(60,120){\circle*{3}}
\put(60,120){\vector(3,-4){12}}

\put(160,170){\makebox(60,20){\large $\langle \hat{S}_1\rangle = 1.75 $}}
\put(160,140){\makebox(60,20){\large $\delta \hat{S}_1^2 = 3.38 $}}

\end{picture}
\caption{\label{prob} Measurement probabilities for a five photon
wedge state at phase shifts of (a) $\Phi=0.288$ and (b) $\Phi=0.623$.
The corresponding averages and variances of the photon number
differences $\hat{S}_1$ are given in the upper
right hand corners. The sketches to the left of the graphs
illustrate the quantized levels with the
highest measurement probabilities in the $S_1$-$S_2$ plane of the
Poincare sphere. The arrows represent the average Stokes vector of
the rotated state.}
\end{figure}

The well defined polarization direction is particularly
visible at phase angles of $\Phi=0.288$ (or $16.5^\circ$),
where the probability of measuring $\mid 3;2 \rangle$ has
its maximal value of 75.2\%, and at $\Phi=0.623$
(or $35.7^\circ$), where the probabilities of measuring
$\mid 2;3 \rangle$ and $\mid 1;4 \rangle$ are both equal
to 42.5\%. Fig. \ref{prob} shows these
two probability distributions, along with a schematic illustration
of the corresponding quantized levels on the Poincare sphere.
As indicated by fig. \ref{prob} (a), the output photon
number distribution can be "switched" between 75.2 \% $S_1=-1$
($\mid 2;3 \rangle$) at $\Phi=-0.288$ and 75.2 \% $S_1=+1$
($\mid 3;2 \rangle$) at $\Phi=+0.288$. Likewise, fig.
\ref{prob} (b) indicates that the photon number distribution
at $\Phi=0$ can be shifted by exactly one photon with only
15\% of the outcomes scattered to different photon numbers.
Considering the fact that the measurement outcomes are discrete,
this shift in the probability distribution is surprisingly
smooth. Specifically, the high fidelity of the $\mid 3;2 \rangle$
component at $\Phi=+0.288$ shown in fig. \ref{prob} (a) suggests
a "polarization wedge" that is much sharper than the photon number
distribution at $\Phi=0$ \cite{qcorr}. The quantum coherence
between the adjacent photon number states induced by the
post-selected photon subtraction thus permits a
surprisingly high level of control at the single photon level.

Although the detailed calculations presented here only apply to the
specific case of five photons, it should be remembered that the
basic features of the statistics for a general $(2n-1)$-photon
wedge state are defined by the Stokes parameter statistics given in
section \ref{sec:stokes}. Since the uncertainty of $\hat{S}_1$ is
always one, the photon number distributions will be limited to
only a few possible measurement outcomes close to $S_1=\pm 1$
for any number of photons. Specifically, the photon number distribution
at $\langle \hat{S}_1 \rangle \approx 1$ will always be similar to
fig. \ref{prob} (a), and the distribution at
$\langle \hat{S}_1 \rangle \approx 2$ will be similar to fig. \ref{prob} (b).
For high $n$, we can therefore expect a maximal probability of $S_1=+1$
at a phase angle of $\Phi \approx 1/n$, where, according to
eq.(\ref{eq:stats}), $\langle \hat{S}_1 \rangle \approx 1$ and
$\delta S_1^2 \approx 2$.
The "switch" from $S_1=-1$ to $S_1=+1$ is therefore also observable
at higher photon numbers.

\section{Conclusions}

In conclusion, it has been shown that it is possible to
obtain highly phase sensitive $(2n\!-\!1)$-photon wedge
states by photon subtraction from $n$ photon pairs
generated in collinear type-II parametric downconversion.
Since the post-selection condition for the generation
of this state can be higher than 18\% regardless of photon
number, the only limitation in extending this scheme to
high photon numbers is the efficiency of the parametric
downconversion. The generation of five photon wedge states
should therefore be well within reach of present technological
capabilities.
Quantitatively, the phase sensitivity of wedge states is
comparable to that of the recently realized path entangled
states, with the advantage that a phase shift can be related
directly to a shift in the photon number distribution observed
in the output. Specifically, phase shifts at the Heisenberg
limit appear as nearly
deterministic transfers of one photon between the two output
ports. $(2n\!-\!1)$-photon wedge states should therefore be highly
suitable for the determination of phase shifts $\Phi$ of
about $1/n$. The generation of polarization wedge states
by downconversion and photon subtraction thus provides a
simple and effective experimental approach to phase measurements
at the Heisenberg limit with non-classical $N$-photon inputs.

Part of this work has been supported by the JST-CREST
project on quantum information processing.


\begin{thebibliography}{xyz00}

%%--general phase problems

\bibitem{Hol93}
M.J. Holland and K. Burnett, Phys. Rev. Lett. {\bf 71},
1355 (1993).

\bibitem{San95}
B.C. Sanders and G.J. Milburn, Phys. Rev. Lett. {\bf 75}, 2944 (1995).

\bibitem{Ou97}
Z.Y. Ou, Phys. Rev. A {\bf 55}, 2598 (1997).

\bibitem{Ber00}
D.W. Berry and H.M. Wiseman,
Phys. Rev. Lett. {\bf 85}, 5098 (2000).

\bibitem{Sod03}
J. S\"oderholm, G. Bj\"ork, B. Hessmo, and S. Inoue,
Phys. Rev. A {\bf 67}, 053803 (2003).

\bibitem{Wan05}
H. Wang and T. Kobayashi, Phys. Rev. A {\bf 71}, 021802(R)
(2005).

\bibitem{Com05}
J.Combes and H.M. Wiseman,
J. Opt. B: Quantum semiclass. Opt. {\bf 7}, 14 (2005).

%%-lithography and multi-photon interference

\bibitem{Jac95}
J. Jacobson, G. Bj\"ork, I.Chuang, and Y. Yamamoto,
Phys. Rev. Lett. {\bf 74}, 4835 (1995).

\bibitem{Fon99}
E.J.S. Fonseca, C.H. Monken, and S. Padua, Phys. Rev. Lett.
{\bf 82}, 2868 (1999).

\bibitem{Bot00}
A.N. Boto, P. Kok, D.S. Abrams, S.L. Braunstein, C.P. Williams,
and J. P. Dowling, Phys. Rev. Lett. {\bf 85}, 2733 (2000).

\bibitem{Ang01}
M. D`Angelo, M.V. Chekhova, and Y. Shih,
Phys. Rev. Lett. {\bf 87}, 013602 (2001).

%%--atomic clocks

\bibitem{Chu00}
I.L. Chuang, Phys. Rev. Lett. {\bf 85}, 2006 (2000).

\bibitem{Jos00}
R. Jozsa, D.S. Abrams, J.P. Dowling, and C.P. Williams,
Phys. Rev. Lett. {\bf 85}, 2010 (2000).

%%--recent phase estimation issues
\bibitem{Eda02}
K. Edamatsu, R. Shimizu, and T. Itoh,
Phys. Rev. Lett. {\bf 89}, 213601 (2002).

\bibitem{Fiu02}
J. Fiurasek, Phys. Rev. A {\bf 65}, 053818 (2002).

\bibitem{Hof04}
H.F. Hofmann, Phys. Rev. A {\bf 70}, 023812 (2004).

\bibitem{Mit04}
M.W. Mitchell, J.S. Lundeen, and A.M. Steinberger,
Nature (London) {\bf 429}, 161 (2004).

\bibitem{Eis05}
H.S. Eisenberg, J.F. Hodelin, G. Khoury, and D. Bouwmeester,
Phys. Rev. Lett. {\bf 94}, 090502 (2005).

\bibitem{ps}
It should be noted that precise photon counting would
require special detectors such as the ones described
in S.Kim, Y.Yamamoto, and H.H. Hogue, Appl. Phys. Lett.
{\bf 74}, 902 (1999). Without such special technology,
the effects of photon losses are usually avoided by
using downconverted light with $r$ much smaller than one.
The effects of higher photon number components can then be
neglected even if the detectors have limited quantum efficiencies
and cannot resolve photon number. However, this method limits
the output intensity and makes it difficult to achieve
high photon numbers.

\bibitem{note}
It may seem a bit counterintuitive that the subtraction
of a photon causes the remaining photons to have the same
polarization as the photon removed from the state.
However, the effect can be understood quite clearly
if one remembers that photon bunching implies that the
photons preferably have the same diagonal polarization.

\bibitem{Eis04}
A similar robustness of bipartite entanglement against
the photon losses caused by limited detector efficiencies has
recently been investigated in H.S. Eisenberg, G. Khoury,
G.A. Durkin, C. Simon, and D. Bouwmeester, Phys. Rev. Lett.
{\bf 93}, 193901 (2004). This paper also gives an interesting
example of parametric downconversion with $r>1$.

\bibitem{qcorr}
It may be worth noting that the apparent localization of the
"wedge" between the quantized measurement results is
somewhat similar to the correlation between coherence and
half-integer photon number results in observable in
finite resolution measurements, as discussed in H.F. Hofmann,
Phys. Rev. A {\bf 61}, 033815 (2000).

\end{thebibliography}
\end{document}